\documentclass[journal]{IEEEtran}
\usepackage{caption}

\usepackage{multicol}
\usepackage{mathtools}
\usepackage{amsthm}
\usepackage{booktabs}

\usepackage{amsfonts}
\usepackage[bookmarksopen=true]{hyperref}
\usepackage{stfloats}
\usepackage{acro}
\usepackage[noadjust]{cite}
\usepackage{multirow}
\usepackage{float}
\usepackage{bm}
\usepackage{enumitem}
\usepackage{amsmath,amssymb}
\usepackage{graphicx}
\usepackage{amsmath,amsfonts}
\usepackage{algorithmic}
\usepackage{epsfig, algorithm}
\usepackage{epstopdf}
\usepackage{caption}
\usepackage[table,xcdraw]{xcolor}
\usepackage{bbm}
\usepackage[geometry]{ifsym}
\usepackage{array}
\usepackage[utf8]{inputenc}
\usepackage[T1]{fontenc}
\usepackage{pifont}
\usepackage{graphicx}   
\usepackage[compatibility=false]{caption}
\usepackage{subcaption} 
\usepackage{algorithmic}
\usepackage{algorithm}
\usepackage{textcomp}
\usepackage{url}
\usepackage{verbatim}
\usepackage{xcolor}
\usepackage[normalem]{ulem}
\providecolor{added}{rgb}{0,0,1}
\providecolor{deleted}{rgb}{1,0,0}

\def\BibTeX{{\rm B\kern-.05em{\sc i\kern-.025em b}\kern-.08em
		T\kern-.1667em\lower.7ex\hbox{E}\kern-.125emX}}

\DeclareAcronym{OFDM}{
short = OFDM,
long = orthogonal frequency division multiplexing
}

\DeclareAcronym{BC}{
short = BC,
long = backscatter communication
}

\DeclareAcronym{Tx}{
short = Tx,
long = base station
}

\DeclareAcronym{MIMO}{
  short = multiple-input multiple-output,
  long = multiple-input multiple-output
}

\DeclareAcronym{XL-MIMO}{
  short = XL-MIMO,
  long = extra-large multiple-input multiple-output
}

\DeclareAcronym{NF}{
  short = NF,
  long = near-field
}

\DeclareAcronym{FF}{
  short = FF,
  long = far-field
}

\DeclareAcronym{LS}{
  short = LS,
  long = least square
}

\DeclareAcronym{FFT}{
  short = FFT,
  long = fast Fourier transform
}

\DeclareAcronym{IDFT}{
  short = FFT,
  long = inverse discrete fast Fourier transform
}

\DeclareAcronym{PLS}{
short = PLS,
long = physical-layer security
}

\DeclareAcronym{PLA}{
short = PLA,
long = physical-layer authentication
}

\DeclareAcronym{SNR}{
short = SNR,
long  = signal-to-noise ratio
}

\DeclareAcronym{ROC}{
short = ROC,
long  = receiver operating characteristic
}

\DeclareAcronym{SSP}{
short = SSP,
long  = spoofing success probability
}

\DeclareAcronym{IoT}{
short = IoT,
long  = internet of things
}

\DeclareAcronym{A-IoT}{
short = A-IoT,
long  = ambient IoT
}

\DeclareAcronym{BD}{
short = BD,
long = backscatter device 
}

\DeclareAcronym{BER}{
short = BER,
long  = bit error rate
}

\DeclareAcronym{CSI}{
short = CSI,
long = channel state information
}

\DeclareAcronym{3GPP}{
short = 3GPP,
long  = 3rd generation partnership project
}

\DeclareAcronym{AWGN}{
short = AWGN,
long  = additive white Gaussian noise
}

\DeclareAcronym{RF}{
short = RF,
long  = radio-frequency
}

\DeclareAcronym{RMSE}{
short = RMSE,
long  = root-mean square error
}

\DeclareAcronym{AN}{
short = AN,
long  = artificial-noise
}

\DeclareAcronym{PDP}{
short = PDP,
long  = power-delay-profile
}

\DeclareAcronym{EW-DSK}{
short = EW-DSK,
long  = element-wise delay shift keying
}

\DeclareAcronym{VR}{
short = VR,
long  = visibility region
}

\DeclareAcronym{ULA}{
short = ULA,
long  = uniform linear array
}

\DeclareAcronym{LoS}{
short = LoS,
long  = line of sight
}

\DeclareAcronym{NLoS}{
short = NLoS,
long  = non LOS
}
\usepackage{cite}
\begin{document}
\title{PHY Authentication for Ambient IoTs in Near-Field XL-MIMO Backscatter
Communication Systems}

\author{Fikiri Salum Uledi, Hafsa Ahmad, Muhammad Bilal Janjua,~\IEEEmembership{Member,~IEEE}, Cagri Ozgenc Etemoglu, and H\"{u}seyin Arslan,~\IEEEmembership{Fellow,~IEEE}
\thanks{F. S Uledi, Hafsa Ahmad and H. Arslan are with the Department of Electrical and Electronics Engineering, Istanbul Medipol University, Istanbul, 34810, T\"{u}rkiye (email: fikiri.uledi@std.medipol.edu.tr, hafsa.ahmad@std.medipol.edu.tr, huseyinarslan@medipol.edu.tr). F. S. Uledi is also with the University of Dar es Salaam, P.O. Box. 35062 Mlimani Road, Dar es Salaam, Tanzania.}
\thanks{ M. B. Janjua and  C. O. Etemoglu are with Turk Telekom R\&D Department, Istanbul 34660, T\"{u}rkiye (email: bilal.janjua@ieee.org, cagriozgenc.etemoglu@turktelekom.com.tr).}
}

\maketitle
\begin{abstract}
This letter leverages near-field (NF) spherical wave-
fronts to derive geometry-tied physical signatures for backscat-
ter device (BD) authentication in extra-large multiple-input
multiple-output (XL-MIMO) based backscatter communication
(BC) systems. Specifically, we propose a binary element-wise
delay shift keying (EW-DSK) BD modulation scheme exploiting
multi-antenna BD design. In the proposed scheme, a deter-
ministic BD-induced delay introduces bit-dependent shifts in
the power-delay-profile (PDP) while preserving the aperture-
dependent NF curvature. Based on the extracted dominant per-
antenna PDP curvature, a compact delay-curvature signature
extractor and an authentication test are developed to mitigate
impersonation attacks. Simulation results demonstrate the pro-
posed scheme’s performance in terms of the curvature-extraction
capability, authentication accuracy and spoofing success prob-
ability (SSP) under varying spoofer distances, signal-to-noise
ratios (SNRs), and effective bandwidths.
\end{abstract}

\vspace{-0.3cm}
\section{Introduction}
\label{sec:intro}
\IEEEPARstart{T}{HE} introduction of~\ac{A-IoT} devices in \ac{3GPP} Release 19 represents a significant step toward massive ultra-low-power, battery-free connectivity envisioned for 6G. In general, \ac{A-IoT} devices are classified according to their design complexity, energy consumption, and communication type, whereby Device 1 and Device 2a, which communicate through backscattering approach, are referred to as \acp{BD}~\cite{3GPP_standard}.~\acp{BD} communicate by modulating their information signals onto incident~\ac{RF} signals from ambient transmitters under a paradigm, namely~\ac{BC}. Although the passive nature of BDs enables BC to achieve low hardware complexity and reduced energy consumption~\cite{An_Overview_on_BC}, it also imposes stringent computational and energy constraints. As a result, conventional cryptographic security becomes difficult to implement, making \ac{BC} links vulnerable to eavesdropping, replay, and impersonation attacks. These constraints motivate the development of lightweight~\ac{PLS} and~\ac{PLA} solutions tailored to cope with~\acp{BD} design limitations~\cite{yang2024physical}.

Most of the existing~\ac{PLS} approaches for \ac{BC} mainly improve confidentiality by exploiting channel randomness,  \ac{AN} design, and spatial processing without relying on heavyweight cryptography~\cite{li2021physical,yang2016physical}. While effective for secrecy enhancement, such methods do not inherently verify the legitimacy of the carrier source leaving the \ac{BC} system vulnerable to carrier source injection or probing attacks. In parallel, existing \ac{PLA} schemes for \ac{BC} rely on \ac{BD}- or channel-dependent features to authenticate the communicating node. For example, attack-tracking and spatially enhanced authentication (BCAuth) are achieved in~\cite{wang2022bcauth}, while robust mutual authentication for multiple \acp{BD} (AuthScatter) is performed in~\cite{zhang2025authscatter}, whereas mobility-aware handover authentication (HABC) is performed in~\cite{yang2025habc}. Recently, a covariance-fingerprint-based \ac{PLA} for ambient \ac{BC} has also been developed in~\cite{AmbAu2025} for spoofing mitigation. Although effective, this approach neither considers signatures induced by \ac{BD} operation nor \ac{BD} design itself. Thus, it does not provide a controlled way to generate discriminative bit-dependent features for authentication.

As far as \ac{BD} design is concerned, multi-antenna \ac{BD} architectures have been investigated in~\cite{niu2023unified} for various spatial-domain modulation schemes. However, these designs mainly target spatial multiple access rather than the deliberate shaping of propagation-delay features for security purposes, and their inter-element spacing is not explicitly specified. In \ac{XL-MIMO} systems, such spacing can be exploited to influence the \ac{NF} propagation characteristics of the \ac{BD} signal. Consequently, antenna-selection-based \ac{BD} modulation with prescribed inter-element spacing can be employed to generate distinctive propagation signatures for \ac{BC} authentication. A related approach has also shown secrecy enhancement via tag selection in a multi-tag \ac{BC} system~\cite{liu2021secrecy}. {Unlike sparse \ac{NF} arrays, which may suffer from focused grating-lobe effects~\cite{zhou2025sparse}, the present study considers a dense \ac{ULA} with half-wavelength inter antenna spacing at the receiver}. Owing to the limited coverage of~\acp{BD}' transmissions and the increased Rayleigh distance associated with large-aperture \ac{XL-MIMO} systems, most \ac{BD} deployments are more likely to operate in the \ac{NF} region of the receiver. In this regime, the spatial response of the \ac{BD} signal depends jointly on range and angle, while the per-antenna propagation delay, as extracted from the \ac{PDP}, exhibits a quadratic delay-curvature profile across the antenna index~\cite{lu2024tutorial,liu2024near}. \ac{PDP}-curvature features have recently been exploited for signal-driven sensing tasks such as blockage detection and mitigation~\cite{Yahya_2025}. In addition, \ac{NF} propagation features were exploited for \ac{PLA} in~\cite{demirci2024visibility}, where a \ac{VR}-based authentication method was developed, although that work focused on active communication rather than \ac{BC}. {Alternatively, movable antennas (MA) architectures may also be used to provide further flexibility for enhancing the geometry-dependent signature discrimination instead of conventional \ac{XL-MIMO}~\cite{zhou2026ma}}. Motivated by these observations, this letter proposes a lightweight BD modulation and authentication framework that intentionally embeds geometry-tied PDP-curvature signatures into the backscattered signal.
\begin{enumerate}
\item We propose a novel \ac{EW-DSK} \ac{BC} scheme exploiting a multiple-antenna element architecture at the \ac{BD}. Therein, a \ac{BD} equipped with two antenna elements spaced by $\lambda/2$, performs element-wise switching with additional induced delays to transmit its information to the ~\ac{XL-MIMO} receiver.
\item We develop a per-antenna \ac{PDP}-signature extractor from a \ac{NF} spherical wavefront's curvature. The extractor identifies the dominant delay profile through per-antenna \ac{PDP} fitting and integrates it with a false-alarm-controlled hypothesis test for authentication.
\item We provide a compact analytical performance analysis and validate the proposed scheme through simulations. The performance evaluation metrics include the delay-curvature extraction accuracy, \ac{ROC} for authentication, and \ac{SSP}.
\end{enumerate}

\begin{figure}[t]
    \centering
    \includegraphics[width=0.85\linewidth]{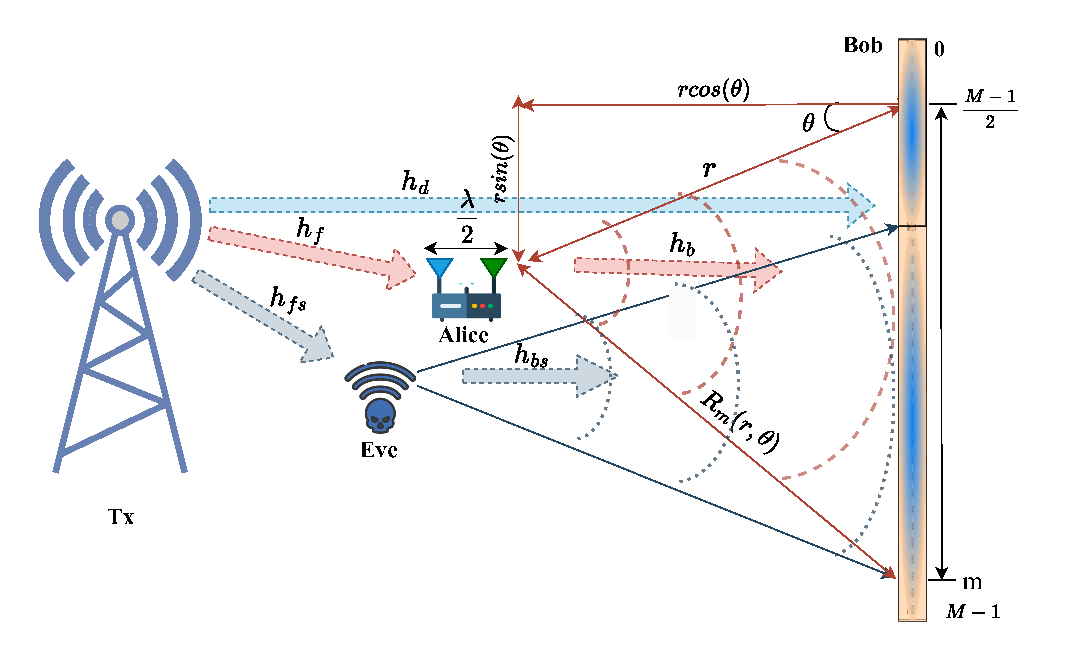}

    \vspace{-0.3cm}
    \caption{System model.}
    \label{fig:image1}
\end{figure}

\vspace{-0.4cm}
\section{Signal and System Model}
\label{sec:sys}
Consider an \ac{OFDM}-based ambient \ac{BC} system in \figurename~\ref{fig:image1}, where \ac{Tx} communicates with an \ac{XL-MIMO} receiver (Bob) over the direct channel $h_d$, while a passive multi-antenna \ac{BD} (Alice) conveys information by reflecting the incident \ac{Tx} signal through the backscatter channel $h_b$ towards Bob. A clone-reflector spoofing attacker (Eve) attempts to deceive Bob by reflecting the same \ac{Tx} signal through a spoofing channel $h_{bs}$ while trying to mimic Alice’s signature. {The scenario is assumed to operate under quasi-static channel conditions, such that the relative geometry remains unchanged over the enrollment and authentication intervals where the small mobility induced curvature displacements cause negligible error that can be absorbed into the aggregate uncertainty induced by curvature extraction}. {Bob is equipped with a \ac{ULA} of $M$ antennas with inter-element spacing $\psi = \lambda/2$}. Alice is assumed to operate in the radiating \ac{NF} of Bob, i.e., at a distance comparable to or smaller than the Rayleigh distance $2D^2/\lambda$, where $\lambda$ denotes wavelength while $D=(M-1)\psi$ represents the array aperture.

\vspace{-0.4cm}
\subsection{Received Signal Model}
Let $x(t)$ denote the complex baseband representation of the waveform sample during one observation interval over the array, the received signal at Bob is modeled as
\begin{equation}\small
\mathbf y(t)=\mathbf h_d x(t)+\alpha \mathbf h_c x(t)+\mathbf w(t),
\label{eq:rx_ext}
\end{equation}
where $\mathbf{h_d}$ is the direct Tx-Bob channel, $\mathbf h_c$ is the cascaded Tx-Alice-Bob channel, $\alpha$ is the Alice's reflection coefficient and $\mathbf w \sim \mathcal{CN}(0,I_M\sigma^2 )$ is \ac{AWGN}, typically modeled as circularly symmetric complex Gaussian noise. The cascaded channel is written as $\mathbf h_c=\beta \mathbf a(r,\theta)$, in which $\beta$ captures the effective complex cascaded propagation gain, accounting for the large-scale attenuation and common propagation phase of the Tx-BD-Bob path, while $ \mathbf a(r,\theta)=\frac{1}{\sqrt{M}}[e^{-j\frac{2\pi}{\lambda}R_0(r,\theta)},e^{-j\frac{2\pi}{\lambda}R_1(r,\theta)},
\ldots,e^{-j\frac{2\pi}{\lambda}R_{M-1}(r,\theta)}]^T$ is the \ac{NF} array response vector associated with \ac{BD} range $r$ and angle $\theta$. For analytical tractability, we represent $\mathbf h_c = \mathbf h_{f}*\mathbf h_{b}$ while  $\mathbf h_{cs} = \mathbf h_{fs}*\mathbf h_{bs}$.

\vspace{-0.3cm}
\subsection{\ac{NF} Geometry and Delay Profile}
From Fig.~\ref{fig:image1}, for antenna index $m=0,\dots,M-1$, let $u_m$ denote the antenna coordinate relative to the array center.
The distance from the \ac{BD} to antenna $m$ is $R_m(r,\theta) = \sqrt{r^2 + u_m^2 - 2 r u_m \sin\theta }$ while the propagation delay is given as $\tau_m = R_m/c$ whose second-order Taylor expansion around $u_m=0$ is expressed as $R_m(r,\theta) \approx r - u_m \sin\theta + \frac{u_m^2}{2r}\cos^2\theta$ with delay across the aperture, given as

\vspace{-0.3cm}
\begin{equation}\small
\tau_m \approx \tau_0 - \frac{u_m}{c}\sin\theta + \frac{u_m^2}{2rc}\cos^2\theta,
\label{eq:taylora}
\end{equation}
where $\tau_0$ denotes the propagation delay associated with the array-center reference point while $c$ represents the speed of light.~\cite{lu2024tutorial,liu2024near}.

\vspace{-0.3cm}
\section{Proposed \ac{EW-DSK} \ac{BD} modulation scheme}
\label{sec:proposed_method}
The proposed scheme involves a multi-antenna \ac{BD} equipped with two antenna elements spaced at $\lambda/2$ from each other, in which \ac{BD} conveys its information through binary element switching combined with additional deterministic \ac{BD}-induced delay. Specifically, bit `0' is transmitted by activating element 1 without additional delay, whereas bit `1' is transmitted by activating element 2 with an imposed delay $\delta_b$. This can be expressed mathematically as

\vspace{-0.3cm}
\begin{equation}\small
\mathcal{M}_b \triangleq
\begin{cases}
0, \quad \mathcal{A}_b=e_i, & b=0\\
\delta_b, \quad \mathcal{A}_b=e_{i+1}, & b=1
\end{cases}~,
\label{eq:mod_rule}
\end{equation}
where $e_i$ and $e_{i+1}$ with $i \in \mathbb{Z}^+$ denote the two selected \ac{BD} antenna elements satisfying an inter-element spacing of $\lambda/2$, $\mathcal{A}_b$ denotes the active \ac{BD} element for bit state $b$, and $\delta_b$ is the imposed deterministic delay associated with bit `1'. For the $m$-th antenna of Bob, the cascaded geometric delay is given as

\vspace{-0.4cm}
\begin{equation}\small
\tau_{b,m}=\frac{r_{{\rm tx},b}+r_{{\rm rx},b,m}}{c},
\label{eq:tau_bm_clear}
\end{equation}
where $r_{{\rm tx},b}$ is the distance between \ac{Tx} and Alice, while $r_{{\rm rx},b,m}$ is the distance from Alice to the $m$-th antenna of Bob. Hence, the effective delay observed at antenna $m$ is $\widetilde{\tau}_{b,m}=\tau_{b,m}+\mathcal{M}_b$. Consequently, Alice's component associated with bit $b$ is expressed as

\vspace{-0.4cm}
\begin{equation}\small
z_{b,m}(t)=\alpha \beta_{b,m}x\left(t-\widetilde{\tau}_{b,m}\right)
e^{-j2\pi f_c \widetilde{\tau}_{b,m}},
\label{eq:zbm}
\end{equation}
where $f_c$ denotes the carrier frequency and $\beta_{b,m}$ captures the cascaded attenuation term. Then, the total signal received at antenna $m$ is given by

\vspace{-0.4cm}
\begin{equation}\small
y_m(t)=h_{d,m}\,x(t-\tau_{d,m}) + z_{b,m}(t)+w_m(t),
\label{eq:ym_final}
\end{equation}
where $\tau_{d,m}$ is the propagation delay of $h_{d,m}$, while $w_m(t)\sim\mathcal{CN}(0,\sigma_w^2)$ denotes \ac{AWGN} at the $m$-th antenna of Bob. In practice, the strong direct-link term $h_{d,m}\,x(t-\tau_{d,m})$ is first mitigated from $y_m(t)$ using prior suppression methods such as~\cite{DSK}, after which the remaining signal is used for per-antenna \ac{PDP} estimation and extraction of the proposed delay-curvature signature. {For analytical tractability, the proposed scheme is developed under under the assumption of fully suppressed direct-link interference. However, in practice, residual leakage may perturb the dominant \ac{PDP} peak, increase delay-curvature estimation error, and is therefore treated here as a practical impairment left for future investigation}.

The selection of $\mathbf{e}_i$ or $\mathbf{e}_{i+1}$ in~\eqref{eq:mod_rule} changes the effective scattering reference point of the \ac{BD}, and consequently modifies the resulting delay-curvature profile extracted from the \ac{PDP}. The imposed delay, $\delta_b$ is introduced as a general design parameter and constrained to ensure delay-state resolvability while preserving the \ac{NF} propagation characteristics. Specifically, $\delta_b$ is constrained as $\frac{\kappa}{B_{\mathrm{eff}}} \le \delta_b \le \xi\,{\Delta \tau}_{\mathrm{NF}}$
where $\kappa \ge 1$ is a safety factor accounting for practical uncertainty in per-antenna \ac{PDP} estimation. $B_{\mathrm{eff}}$ is the effective signal bandwidth, and $\xi\in(0,1)$ is a design margin that limits $\delta_b$ to a fraction of the intrinsic \ac{NF} delay spread, defined as ${\Delta \tau}_{\mathrm{NF}} = \max_m \tau_{b,m} - \min_m \tau_{b,m}$. The lower bound guarantees delay-state resolvability, whereas the upper bound preserves the aperture-dependent \ac{NF} curvature signature. The influence of $\delta_b$ on bit-state resolvability is quantified by normalized delay-resolution index $\rho_{\mathrm{res}}\triangleq \delta_b B_{\mathrm{eff}}$. Over a given observation, the delay-state separability is then defined as $\mathcal{J}_{\mathrm{sep}} \triangleq \frac{|\mu_1-\mu_0|}{\sqrt{\sigma_0^2+\sigma_1^2}}$, where $\mu_0$,$\sigma_0^2$ and $\mu_1$,$\sigma_1^2$ are the corresponding sample mean and variance for bit state `0' and `1', respectively. A larger $\mathcal{J}_{\mathrm{sep}}$ indicates stronger discrimination between the two imposed delay states of the curvatures.

\vspace{-0.4cm}
\subsection{Delay-Curvature Signature Extraction}
\label{subsec:signature}
From $z_{b,m}(t)$ in~\eqref{eq:zbm}, Bob estimates the \ac{PDP} curvature $P_{b,m}(\tau)$ and extracts the dominant delayed curvature, such that $\widehat{\tau}_{b,m}=\arg\max_{\tau}P_m(\tau)$, which follows a quadratic \ac{NF} propagation characteristic throughout the aperture while including $\delta_b$. Accordingly, Bob fits the model as
\begin{equation}\small
\widehat{\tau}_{b,m}\approx d_0+d_1u_m+d_2u_m^2+\delta_b+\nu_m,
\label{eq:delay_quad_rect}
\end{equation}
where $d_0$ is the common delay term, $d_1$ is the linear delay-slope coefficient across the aperture, and $d_2$ is the quadratic coefficient that captures the \ac{NF} curvature effect while $\nu_m$ captures curvature-delay estimation error. Based on \eqref{eq:delay_quad_rect}, Bob forms the compact delay-curvature signature $\mathbf s=[\widehat d_1,\widehat d_2]^T$, where $\widehat d_1$ and $\widehat d_2$ denote the fitted linear and quadratic coefficients, respectively. {Computational overhead is incurred at Bob, while the \ac{BD} remains passive and low-complexity. The complexity is dominated by FFT-based per-antenna \ac{PDP} extraction, requiring $\mathcal{O}(MN\log N)$ over $M$ antennas and $N$ subcarriers, followed by delay-bin search and quadratic fitting with complexities $\mathcal{O}(MN)$ and $\mathcal{O}(M)$, respectively. Thus, the overall complexity is $\mathcal{O}(MN\log N+MN+M)$.}

\vspace{-0.4cm}
\subsection{Authentication Decision Rule}
\label{subsec:auth}
During an enrollment phase, Bob stores the legitimate reference signatures $\{\mathbf s^{(\mathrm{ref})}_0,\mathbf s^{(\mathrm{ref})}_1\}$ corresponding to the $\delta_b$ such that, for each new observation, Bob computes

\vspace{-0.3cm}
\begin{equation}\small
\eta=\min_{b\in\{0,1\}}\left\|\mathbf s^{(\mathrm{obs})}-\mathbf s^{(\mathrm{ref})}_b\right\|_2,
\label{eq:eta_ext}
\end{equation}
and performs the binary hypothesis test such as

\vspace{-0.3cm}
\begin{equation}\small
\mathcal H_a:\eta\le \eta_{\mathrm{th}},\qquad
\mathcal H_e:\eta> \eta_{\mathrm{th}},
\label{eq:test_ext}
\end{equation}
where $\mathcal H_a$ denotes the legitimate-tag (Alice) hypothesis, $\mathcal H_e$ denotes the illegitimate-tag (Eve) hypothesis while $\eta_{\mathrm{th}}$ is the threshold selected to satisfy the target false-alarm probability $P_{\mathrm{FA}}=\Pr(\eta>\eta_{\mathrm{th}}\mid \mathcal H_a)$,
{where its formulation is done under quasi-static channel assumption to ensure that the enrolled signature remains stable over the authentication interval.}

\vspace{-0.3cm}
\section{Performance Analysis}
\label{sec:analysis}
This section develops a compact analytical characterization of the proposed scheme by examining how \ac{PDP} curvature-estimation error propagates into delay-curvature signature estimation and how the resulting signature uncertainty impacts threshold selection and spoofing-detection performance.

\vspace{-0.4cm}
\subsection{Curvature-Extraction Error and Signature Estimation}
\label{subsec:curvature_error}
{The proposed scheme is analyzed under dominant backscatter-path link where additional reflected backscatter components are expected to remain substantially weaker than the principal backscattered component due to the extra cascaded attenuation.} From the fitted model in \eqref{eq:delay_quad_rect}, the curvature samples extracted across the XL-MIMO aperture are  collected into the vector $ \widehat{\boldsymbol{\tau}}_{b,m} = \big[\widehat{\tau}_b[0],\widehat{\tau}_b[1],\ldots,\widehat{\tau}_b[M-1]\big]^T$ whereby its matrix form representation can then be written as $\widehat{\boldsymbol{\tau}}_{b,m} =\mathbf X \boldsymbol{\Phi}_b + \boldsymbol{\nu},$
where

\vspace{-0.4cm}
\begin{equation}\small
\mathbf X=
\begin{bmatrix}\small
1 & u_0 & u_0^2\\
1 & u_1 & u_1^2\\
\vdots & \vdots & \vdots\\
1 & u_{M-1} & u_{M-1}^2
\end{bmatrix},
\qquad
\boldsymbol{\Phi}_b=
\begin{bmatrix}
d_0+\delta_b\\
d_1\\
d_2
\end{bmatrix},
\label{eq:X_theta}
\end{equation}
while $\boldsymbol{\nu}=[\nu_0,\nu_1,\ldots,\nu_{M-1}]^T$. The \ac{LS} estimate of the parameter vector is then given as $\widehat{\boldsymbol{\Phi}}_b =(\mathbf X^T\mathbf X)^{-1}\mathbf X^T\widehat{\boldsymbol{\tau}}_{b,m}$. Since the authentication feature is formed by the linear and quadratic curvature terms, we define

\vspace{-0.3cm}
\begin{equation}\small
\widehat{\mathbf s}_b
=
\mathbf J\widehat{\boldsymbol{\Phi}}_b,
\qquad
\mathbf J=
\begin{bmatrix}
0 & 1 & 0\\
0 & 0 & 1
\end{bmatrix},
\label{eq:sig_from_theta}
\end{equation}

\vspace{-0.3cm}
whereby $\widehat{\mathbf s}_b=[\widehat d_1,\widehat d_2]^T$. Assuming that the curvature-estimation errors are zero-mean with covariance $\mathbb E[\boldsymbol{\nu}\boldsymbol{\nu}^T]=\sigma_\nu^2\mathbf I_M$, the covariance of the LS estimator is approximated as
$\mathrm{Cov}(\widehat{\boldsymbol{\Phi}}_b) \approx \sigma_\nu^2(\mathbf X^T\mathbf X)^{-1}$ and consequently $\mathrm{Cov}(\widehat{\mathbf s}_b)
\approx \sigma_\nu^2 \mathbf J(\mathbf X^T\mathbf X)^{-1}\mathbf J^T.$
{The given model is adopted as a tractable first-order approximation. More generally, practical curvature-extraction errors satisfy non nonzero bias, $\mathbb{E}\{\boldsymbol{\nu}\}=\boldsymbol{\mu}_{\nu}$ and spatially correlated covariance, $\mathrm{Cov}(\boldsymbol{\nu})=\boldsymbol{\Sigma}_{\nu}$, due to residual interference, synchronization mismatch and hardware impairments. Modeling such impairment-dependent error structures is left for future extension.} Under this approximation, the mean-square signature error satisfies

\vspace{-0.3cm}
\begin{equation}\small
\mathbb{E}\!\left[\left\|\hat{\mathbf{s}}_{b}-\mathbf{s}_{b}\right\|_2^2\right]
\approx \sigma_{\nu}^{2}\, \operatorname{tr}\!\left(\mathbf{J}\left(\mathbf{X}^{T}\mathbf{X}\right)^{-1}\mathbf{J}^{T}\right),
\label{eq:theta_cov1}
\end{equation}

\vspace{-0.2cm}
showing that the accuracy of the extracted delay-curvature signature depends on both curvature-extraction quality and aperture geometry. In particular, reducing $\sigma_\nu^2$ through more reliable per-antenna \ac{PDP} estimation, higher \ac{SNR}, or larger $B_{\mathrm{eff}}$ improves curvature estimation, while a larger aperture lowers the estimation variance through better conditioning of $\mathbf X^T\mathbf X$. {Increasing the number of antennas enlarges the effective aperture and strengthens the \ac{NF} delay-curvature signature, with a weaker effect toward the \ac{FF} regime~\cite{Yahya_2025}.}

\begin{figure*}[t]
  \centering
  \begin{minipage}[t]{0.3\linewidth}
    \raggedright
    \includegraphics[width=\linewidth]{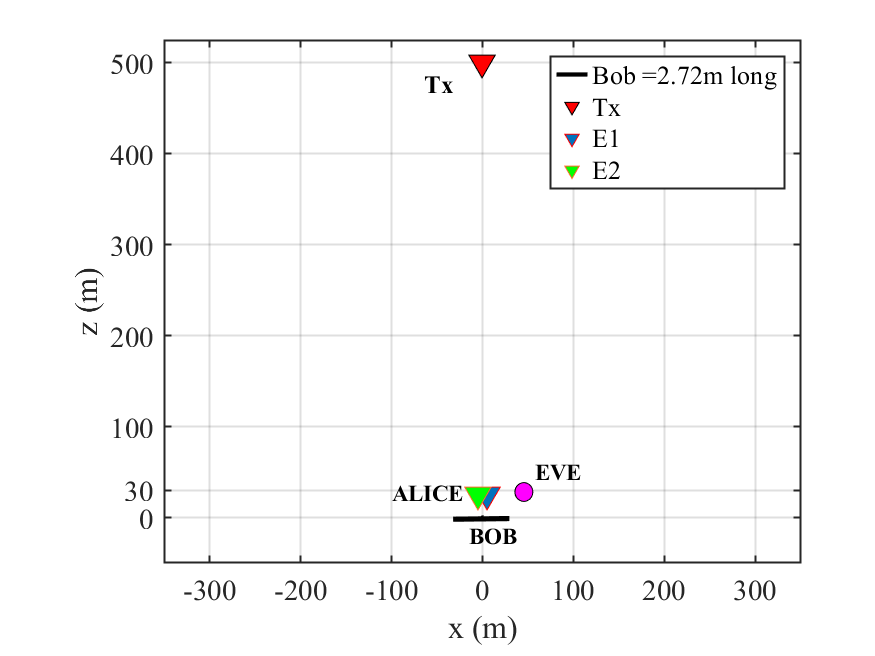}
    
    \vspace{-0.2cm}
    \caption{System geometry.}
     \label{fig:System_Geometr}
  \end{minipage}\hfill
  \begin{minipage}[t]{0.3\linewidth}
    \centering
    \includegraphics[width=\linewidth]{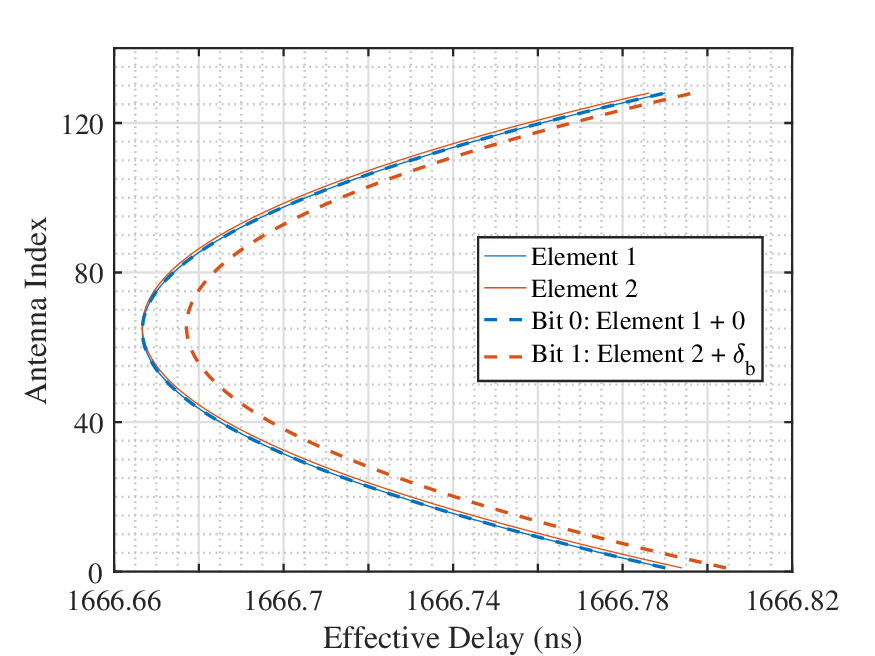}

    \vspace{-0.2cm}
    \caption{\ac{PDP} curvature for the bit states.}
    \label{fig:PDP}
  \end{minipage}\hfill
  \begin{minipage}[t]{0.3\linewidth}
    \raggedleft
    \includegraphics[width=\linewidth]{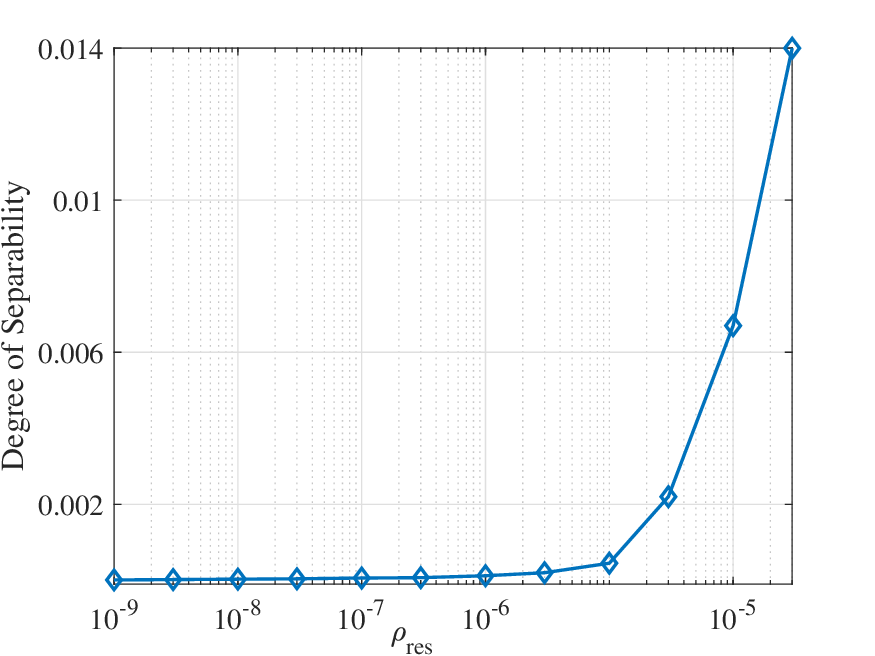}
    
    \vspace{-0.2cm}
    \caption{Curvature separability index.}
     \label{fig:Tau_plot}
  \end{minipage}
\end{figure*}

\vspace{-0.5cm}
\subsection{Statistical Threshold Behavior and Spoofing Analysis}
\label{subsec:auth_margin}
Building on the decision rule in \eqref{eq:eta_ext}, let $\mathbf e_s =\mathbf s^{(\mathrm{obs})} - \mathbf s_b^{(\mathrm{ref})}$ to denote the signature error relative to the legitimate (Alice) reference for bit state $b$, then under the $\mathcal H_a$, the observed signature can be expressed as $\mathbf s^{(\mathrm{obs})} = \mathbf s_b^{(\mathrm{ref})} + \mathbf e_s$.
For analytical tractability, assume $\mathbf e_s \sim \mathcal N(\mathbf 0,\sigma_s^2\mathbf I_2)$, where $\sigma_s^2$ captures the aggregate uncertainty induced by curvature extraction, curve fitting, and residual cancellation error. Then the decision statistic 
$\eta=\|\mathbf e_s\|_2,$ in \eqref{eq:eta_ext}, follows a Rayleigh distribution, resulting in

\vspace{-0.3cm}
\begin{equation}\small
P_{\mathrm{FA}} = \Pr(\eta>\eta_{\mathrm{th}} \mid \mathcal H_a) =\exp\!\left(-\frac{\eta_{\mathrm{th}}^2}{2\sigma_s^2}\right).
\label{eq:pfa_rayleigh}
\end{equation}
Hence, for a target $P_{\mathrm{FA}}^*$, the threshold can be selected as $ \eta_{\mathrm{th}} = \sigma_s\sqrt{-2\ln(P_{\mathrm{FA}}^*)}$. Under the $\mathcal H_e$, let the Eve-induced signature be denoted by $\mathbf s_{\mathrm{sp}}$, such that $
\mathbf s^{(\mathrm{obs})} = \mathbf s_{\mathrm{sp}} + \mathbf e_s$, then the geometry-induced signature separation is obtained as $\Delta_s= \min_{b\in\{0,1\}}\left\|\mathbf s_{\mathrm{sp}}-\mathbf s_b^{(\mathrm{ref})}\right\|_2$. In this case, the statistic $\eta$ follows a Rician distribution with noncentrality parameter $\Delta_s$, and the spoof detection probability is given as

\vspace{-0.3cm}
\begin{equation}\small
P_D=\Pr(\eta>\eta_{\mathrm{th}} \mid \mathcal H_e)=Q_1\!\left(\frac{\Delta_s}{\sigma_s},\frac{\eta_{\mathrm{th}}}{\sigma_s}\right),
\label{eq:pd_marcum}
\end{equation}
where $Q_1(\cdot,\cdot)$ is the first-order Marcum-$Q$ function. Equivalently, the \ac{SSP} is obtained as

\vspace{-0.3cm}
\begin{equation}\small
P_{\mathrm{sp}}=1-P_D=\Pr(\eta\le\eta_{\mathrm{th}} \mid \mathcal H_e).
\label{eq:pspoof}
\end{equation}

\eqref{eq:pd_marcum}--\eqref{eq:pspoof} highlight the fundamental tradeoff of the proposed scheme where authentication improves when $\Delta_s$ becomes larger and the estimation uncertainty $\sigma_s^2$ becomes smaller.

\begin{figure*}[t]
  \centering
    \begin{minipage}[t]{0.3\linewidth}
    \raggedright
    \includegraphics[width=\linewidth]{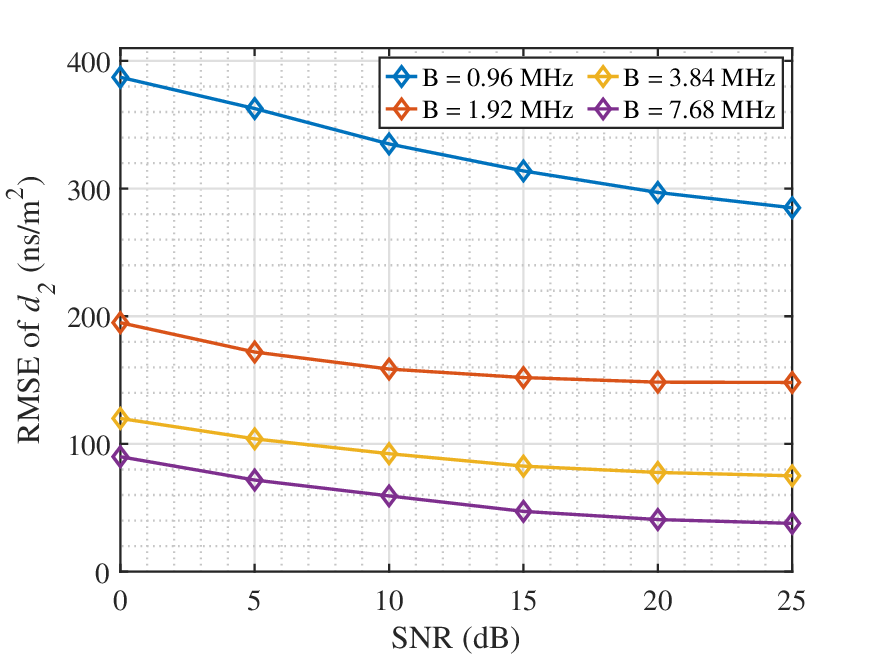}

    \vspace{-0.25cm}
    \caption{Curvature-estimation RMSE}
     \label{fig:RMSE}
  \end{minipage}\hfill
  \begin{minipage}[t]{0.3\linewidth}
    \centering
    \includegraphics[width=\linewidth]{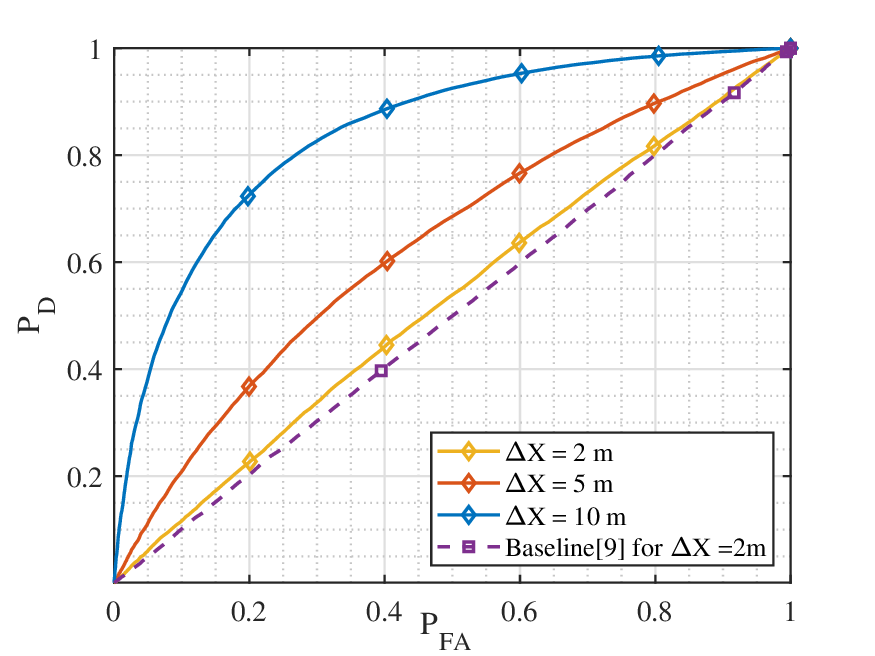}
    
    \vspace{-0.25cm}
    \caption{ROC for Eve offsets.}
    \label{fig:ROC}
  \end{minipage}\hfill
  \begin{minipage}[t]{0.3\linewidth}
    \raggedleft
    \includegraphics[width=\linewidth]{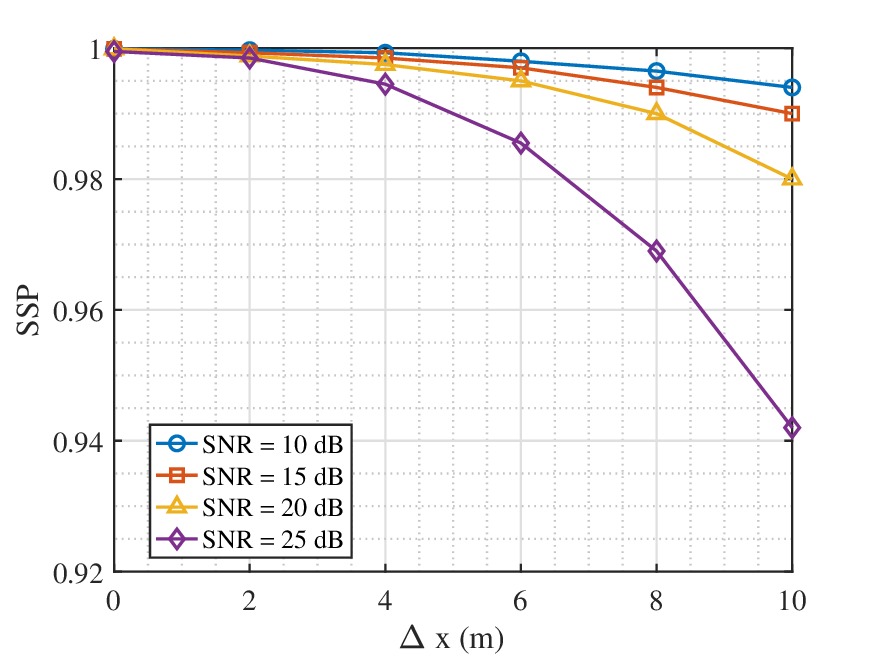}

    \vspace{-0.25cm}
    \caption{SSP versus Eve offset.}
     \label{fig:SSP}
  \end{minipage}
\end{figure*}

\vspace{-0.2cm}
\begin{table}[h!]
    \centering
    \scriptsize

 \vspace{-0.2cm}
  \caption{Key simulation parameters.}
  \label{tab:sim_param}
  \footnotesize
  \setlength{\tabcolsep}{6pt}
  \renewcommand{\arraystretch}{0.8}
  \setlength{\extrarowheight}{1pt}
  \begin{tabular}{|l|c|}
    \hline
    \textbf{Parameter} & \textbf{Value} \\
    \hline
    Central frequency , $f_c$ & $7~\mathrm{GHz}$ \\
    
    Bob's number of antennas, $M$ & $128$ , centred at x(m) = $0$ m \\
    
   Alice's inter-element spacing & $\lambda/2$ \\
    
    DFT size, $N$ & $64$ \\
    
    Subcarrier spacing, $\Delta$f  & $15~\mathrm{kHz}$ \\
    
    $B_{\mathrm{eff}}$ = $N$$\Delta$f & $960~\mathrm{kHz}$ \\
    
    Tx's coordinates & x(m) = $0$m, z(m) = $500$m \\
    
    Alice's coordinates & x(m) = $0$ m, z(m) = $25$ m \\
    
    Eve's coordinates & x(m) = \{0:2:10\}m, z(m) = $30$m \\
    
    Alice's reflection coefficient, $\alpha$ & $1$ \\
    
    Induced delay, $\delta_b$  & $10~\mu\text{s}$ \\
    \hline
  \end{tabular}
\end{table}

\vspace{-0.85cm}
\section{Simulation Results}
\label{sec:sims}
This section presents a numerical evaluation of the proposed scheme. The scenario is graphically illustrated in \figurename~\ref{fig:System_Geometr} while the key simulation parameters are given in Table~\eqref{tab:sim_param}. The results obtained are compared with the baseline~\cite{AmbAu2025} set in the same \ac{NF} geometry as the proposed scheme.

\figurename~\ref{fig:PDP} illustrates the per-antenna \ac{PDP} curvature associated with the two bit states. The extracted profiles preserve aperture-dependent \ac{NF} propagation characteristics, confirming that the received \ac{BD} signal remains governed by the underlying propagation of spherical wave fronts. Moreover, the induced-delay-state exhibits a clear shift relative to the non-delayed one, which confirms that $\delta_b$ improves bit-state discrimination through a controllable curvature resolution.

In \figurename~\ref{fig:Tau_plot}, the curvature separability is examined versus the normalized delay-resolution $\rho_{\mathrm{res}}$. For small $\rho_{\mathrm{res}}$, the separability remains limited, indicating poor curvature resolvability. As $\rho_{\mathrm{res}}$ increases, separability improves monotonically, with a more pronounced rise at higher values. This confirms that increasing $\delta_b$ enhances discrimination between the two \ac{BD} bit states by shifting the effective scattering reference point.

\figurename~\ref{fig:RMSE} shows curvature-estimation \ac{RMSE} versus \ac{SNR} for different values of $B_{\mathrm{eff}}$. \ac{RMSE} is observed to decrease with increasing \ac{SNR} for all $B_{\mathrm{eff}}$, indicating that the proposed curvature signature becomes more reliable as observation quality improves due to reduced estimation error. In addition, increasing $B_{\mathrm{eff}}$ consistently reduces \ac{RMSE}, because a wider frequency range provides a finer delay resolution and therefore improves the quality of the fitted curvature parameters.

\figurename~\ref{fig:ROC} presents \ac{ROC} of the proposed detector for different Eve's displacement offsets. It is observed that when Eve is nearly co-located with Alice, e.g. for $\Delta x = 2$ m, the signature separation is negligible, and \ac{ROC} curve approaches the diagonal with $P_D \approx P_{FA}$. However, as $\Delta x$ increases to $5$ m and $10$ m, the signature becomes more distinguishable, producing a clearly improved \ac{ROC}. This demonstrates that the geometry-dependent features exploited by the proposed \ac{PLA} improve detection for attackers located farther from Alice, which is consistent with practical spoofing scenarios. In addition, the results are compared to the baseline~\cite{AmbAu2025}. As shown, for the same $\Delta x$, the proposed scheme achieves a more favorable \ac{ROC} than the baseline~\cite{AmbAu2025}, hence highlighting the effectiveness of the proposed \ac{PLA}.

\figurename~\ref{fig:SSP} shows the \ac{SSP} versus Eve’s displacement offset $\Delta x$ for different \ac{SNR} values. It is observed that, for all  \acp{SNR}, the \ac{SSP} remains close to unity when Eve is located very close to Alice, indicating that a nearly co-located attacker can induce highly similar signatures and is therefore difficult to distinguish from the legitimate node. As $\Delta x$ increases, the \ac{SSP} decreases monotonically, showing that Eve becomes progressively less capable of reproducing Alice’s signature. In addition, higher \ac{SNR} leads to a faster reduction in \ac{SSP}, since more reliable delay-curvature estimation enables the receiver to exploit the geometry-induced signature mismatch more effectively. This behavior confirms that the proposed authentication mechanism is fundamentally geometry-driven: as Eve moves away from Alice, the induced signature departs from the enrolled reference, thereby reducing the likelihood of false acceptance, while higher \ac{SNR} further strengthens this discrimination capability.

\vspace{-0.4cm}
\section{Conclusion}
\label{sec:conclusion}
This letter proposes a \ac{PLA} approach for~\ac{A-IoT} in \ac{NF} \ac{XL-MIMO} \ac{BC} systems employing a binary \ac{EW-DSK} \ac{BD} modulation scheme. In the proposed scheme, \ac{BD} embeds geometry-tied delay-curvature signatures in the backscattered signal at the \ac{XL-MIMO} receiver to enable lightweight \ac{BD} authentication. The main contributing factors in authentication include per-antenna \ac{PDP} extraction, quadratic fitting, and false-alarm-controlled hypothesis testing. The analytical and simulation results demonstrate improved spoofing resistance with reliable delay-curvature signature estimation, with further gains achieved through higher \acp{SNR} and wider effective bandwidth. Future work will investigate robustness under mobility and time-varying channels, together with the joint optimization of antenna-element spacing and delay-offset design.

\vspace{-0.5cm}
\section*{Acknowledgment}
This work is supported by The Scientific and Technological Research Council of Türkiye (TÜBİTAK) 1515 Frontier R\&D Laboratories Support Program for Türk
Telekom 6G R\&D Lab under project number 5249902. 

\vspace{-0.3cm}
\bibliography{IEEEfull,ULEDI_WCL2026-2125_references.bib}
\bibliographystyle{IEEEtran}
\end{document}